\def\Im{{\text{Im}}\,}

\def\kF{k_{\text{F}}}
\def\vF{v_{\text{F}}}
\def\NF{N_{\text{F}}}
\def\me{m_{\text{e}}}

\def\sgn{{\text{sgn\,}}}
\def\be{\begin{equation}}
\def\ee{\end{equation}}
\def\bea{\begin{eqnarray}}
\def\eea{\end{eqnarray}}
\def\bse{\begin{subequations}}
\def\ese{\end{subequations}}

\documentclass[prb,twocolumn,showpacs,amsmath,amssymb,eqsecnum,preprintnumbers]{revtex4}
\usepackage{graphicx}
\usepackage{dcolumn}
\usepackage{bm}
\begin{document}
\title{Theory of Helimagnons in Itinerant Quantum Systems III: \\
       Quasiparticle Description}
\author{T. R. Kirkpatrick$^1$, D. Belitz$^{2,3}$ and Ronojoy Saha$^2$}
\affiliation{$^{1}$ Institute for Physical Science and Technology,
                    and Department of Physics, University of Maryland, College Park,
                    MD 20742, USA\\
             $^{2}$ Department of Physics and Institute of Theoretical Science,
                    University of Oregon, Eugene, OR 97403, USA\\
             $^{3}$ Materials Science Institute, University of Oregon, Eugene,
                    OR 97403, USA}
\date{\today}

\begin{abstract}
In two previous papers we studied the problem of electronic properties in a
system with long-ranged helimagnetic order caused by itinerant electrons. A
standard many-fermion formalism was used. The calculations were quite tedious
because different spin projections were coupled in the action, and because of
the inhomogeneous nature of a system with long-ranged helimagnetic order. Here
we introduce a canonical transformation that diagonalizes the action in spin
space, and maps the problem onto a homogeneous fermion problem. This
transformation to quasiparticle degrees of freedom greatly simplifies the
calculations. We use the quasiparticle action to calculate single-particle
properties, in particular the single-particle relaxation rate. We first
reproduce our previous results for clean systems in a simpler fashion, and then
study the much more complicated problem of three-dimensional itinerant
helimagnets in the presence of an elastic relaxation rate $1/\tau$ due to
nonmagnetic quenched disorder. Our most important result involves the
temperature dependence of  the single-particle relaxation rate in the ballistic
limit, $\tau^2 T\epsilon_{\text{F}} > 1$, for which we find a linear
temperature dependence. We show how this result is related to a similar result
found in nonmagnetic {\em two-dimensional} disordered metals.

\end{abstract}


\maketitle

\section{Introduction}
\label{sec:I}

In two previous papers, hereafter denoted by I and
II,\cite{Belitz_Kirkpatrick_Rosch_2006a, Belitz_Kirkpatrick_Rosch_2006b} we
considered various properties of clean itinerant helimagnets in their ordered
phase at low temperatures. These papers considered and discussed in some detail
the nature of the ordered state, and in particular the Goldstone mode that
results from the spontaneously broken symmetry. We also calculated a variety of
electronic properties in the ordered state that are influenced by the Goldstone
mode or helimagnon, which physically amounts to fluctuations of the helical
magnetization. For various observables, we found that couplings between
electronic degrees of freedom and helimagnon fluctuations lead to a nonanalytic
(i.e., non-Fermi-liquid-like) temperature dependence at low temperature. For
most quantities, this takes the form of corrections to Fermi-liquid behavior,
but in some cases, e.g., for the single-particle relaxation rate, the
nonanalytic dependence constitutes the leading low-temperature behavior.

A prototypical itinerant helimagnet is MnSi. At low temperatures and ambient
pressure, the ground state of MnSi has helical or spiral order, where the
magnetization is ferromagnetically ordered in the planes perpendicular to some
direction ${\bm q}$, with a helical modulation of wavelength $2\pi /\vert {\bm
q}\vert$ along the ${\bm q}$ axis.\cite{Ishikawa_et_al_1976} MnSi displays
helical order below a temperature $T_c\approx 30\,\text{K}$ at ambient
pressure, with $2\pi /\vert{\bm q}\vert \approx 180\,\text{\AA}$. That is, the
pitch length scale is large compared to microscopic length scales. Application
of hydrostatic pressure $p$ suppresses $T_c$, which goes to zero at a critical
pressure $p_{\,c}\approx 14\,{\text{kbar}}$.\cite{Pfleiderer_et_al_1997}
Physically, the helimagnetism is caused by the spin-orbit interaction, which
breaks lattice inversion symmetry. In a Landau theory this effect leads to a
term of the form ${\bm m}\cdot ({\bm\nabla}\times{\bm m})$ in the Landau free
energy, with ${\bm m}$ the local magnetization, and a prefactor proportional to
$\vert{\bm q}\vert$.\cite{Dzyaloshinski_1958, Moriya_1960}

In I and II we used a technical description based on itinerant electrons
subject to an effective inhomogeneous external field that represents
helimagnetic order, with helimagnon fluctuations coupled to the remaining
electronic degrees of freedom. We emphasize that for the theory developed in
either papers I and II, or in the current paper and a forthcoming paper IV, it
is irrelevant whether the helimagnetism is caused by the conduction electrons,
or whether the conduction electrons experience a background of helimagnetic
order caused by electrons in a different band. The Gaussian or
`non-interacting' part of the action was not diagonal in either spin or wave
number space, and the latter property reflected the fact that the system is
inhomogeneous. These features substantially complicated the explicit
calculations performed in II. In order to make progress beyond the discussion
in II, and to discuss the effects of quenched disorder in particular, it
therefore is desirable to find a technically simpler description. In the
current paper, our first main result is the construction of a canonical
transformation that diagonalizes the action in spin space and simultaneously
makes the Gaussian action diagonal in wave number space. The new transformed
action makes our previous calculations much simpler than before. It also
enables us to extend our previous work in a number of ways. In particular, we
will treat the much more complicated problem of the quasiparticle properties of
helimagnets in the presence of non-magnetic quenched disorder.

The study of the electronic properties of disordered metals has produced a
variety of surprises over the past thirty years. The initial work on this
subject was mostly related to diffusive electrons and the phenomena known as
weak-localization and/or Altshuler-Aronov (AA) effects (for reviews see, e.g.,
Refs.\ \onlinecite{Altshuler_Aronov_1984, Lee_Ramakrishnan_1985}). In the clean
limit, mode-mode coupling effects analogous to the AA effects have been shown
to lead to a nonanalytic wave number dependence of the spin susceptibility at
$T=0$.\cite{Belitz_Kirkpatrick_Vojta_1997} More recently, disordered
interacting (via a Coulomb interaction) electron systems have been studied in
the ballistic limit, $T\tau > 1$, but still at temperatures low compared to all
energy scales other than $1/\tau$, with $\tau$ the elastic scattering
rate.\cite{Zala_Narozhny_Aleiner_2001} Interestingly, in this limit it has been
shown that for two-dimensional (2D) systems, the temperature correction to the
elastic scattering rate is proportional to $T$, i.e., shows non-Fermi liquid
behavior. In contrast, in 3D systems the corresponding correction is
proportional to $T^{2}\ln (1/T)$, i.e., the behavior is marginally
Fermi-liquid-like, with a logarithmic
correction.\cite{Sergeev_Reizer_Mitin_2005} The second main result of the
current paper is that in the ballistic limit, the low-temperature correction to
the single-particle relaxation rate in ordered helimagnets is linear in $T$.
The technical reason for why a 3D disordered itinerant helimagnet behaves in
certain ways in close analogy to a 2D nonmagnetic disordered metal will be
discussed in detail below. Transport properties, in particular the electrical
conductivity, will be considered in a separate paper, which we will refer to as
paper IV.\cite{paper_IV}

The organization of this paper is as follows. In Sec.\ \ref{sec:II} we
introduce a canonical transformation that vastly simplifies the calculation of
electronic properties in the helimagnet state. In Sec.\ \ref{sec:III} we
calculate various single-particle and quasiparticle properties at low
temperatures in both clean and disordered helimagnetic systems. In the latter
case we focus on the ballistic limit (which is slightly differently defined
than in nonmagnetic materials), where the various effects are most interesting,
and which is likely of most experimental interest, given the levels of disorder
in the samples used in previous experiments. The paper is concluded in Sec.\
\ref{sec:IV} with a summary and a discussion. Throughout this paper we will
occasionally refer to results obtained in I and II, and will refer to equations
in these papers in the format (I.x.y) and (II.x.y), respectively.

\section{Canonical Transformation to Quasiparticle Degrees of Freedom}
\label{sec:II}

In this section we start with an electron action that takes into account
helical magnetic order and helical magnetic fluctuations. The fundamental
variables in this description are the usual fermionic (i.e., Grassmann-valued)
fields $\bar{\psi}_{\alpha }(x)$ and $\psi_{\alpha}(x)$. Here $x = ({\bm
x},\tau)$ is a four-vector that comprises real space position ${\bm x}$ and
imaginary time $\tau$, and $\alpha$ is the spin index. Due to the helical
magnetic order, the quadratic part of this action is not diagonal in either the
spin indices or in wave number space. We will see that there is a canonical
transformation which leads, in terms of new Grassmann variables, to an action
that is diagonal in both spin and wave number space. This transformed action
enormously simplifies calculations of the electronic properties of both clean
and dirty helical magnetic metals.

\subsection{The action in terms of canonical variables}
\label{subsec:II.A}

In II we derived an effective action for clean itinerant electrons in the
presence of long-range helical magnetic order, and helical magnetic
fluctuations interacting with the electronic degrees of freedom. This action
can be written (see Eq.\ (II.3.13)),
\be
S_{\text{eff}}[\bar\psi,\psi] = S_0[\bar\psi,\psi] + \frac{\Gamma
_{\text{t}}^2}{2}\int dx dy\ \delta
n_{\text{s}}^i(x)\chi_{\text{s}}^{ij}(x,y)\delta n_{\text{s}}^j(y),
\label{eq:2.1}
\ee
where $n_{\text{s}}^i(x) = \bar\psi_{\alpha}(x) \sigma_{\alpha\beta}^i
\psi_{\alpha}(x)$ is the electronic spin density, $\sigma^i$ ($i=1,2,3$)
denotes the Pauli matrices, $\delta n_{\text{s}}^i = n_{\text{s}}^i - \langle
n_{\text{s}}^i\rangle$ is the spin density fluctuation, $\Gamma_{\text{t}}$ is
the spin-triplet interaction amplitude, and $\int dx = \int d{\bm x}
\int_0^{1/T}$. Here and it what follows we use units such that $k_{\text{B}} =
\hbar = 1$. In Eq. (\ref{eq:2.1}), $S_0$ denotes an action,
\bse
\label{eqs:2.2}
\be
S_0[\bar\psi,\psi] = {\tilde S}_0[\bar\psi,\psi] + \int dx\ {\bm H}_0({\bm
x})\cdot {\bm n}_{\text{s}}(x),
\label{eq:2.2a}
\ee
where,
\be
{\bm H}_0({\bm x}) = \Gamma_{\text{t}}  \langle {\bm n}_{\text{s}}(x)\rangle =
\Gamma_{\text{t}}\,{\bm m}(x)
\label{eq:2.2b}
\ee
is proportional to the average magnetization ${\bm m}({\bm x}) = \langle
{\bm n}_{\text{s}}({\bm x})\rangle$. In the helimagnetic state,
\be
{\bm H}_0({\bm x}) = \lambda \bigl(\cos({\bm q}\cdot{\bm x}),\sin ({\bm q}\cdot
{\bm x}), 0\bigr),
\label{eq:2.2c}
\ee
\ese
where ${\bm q}$ is the pitch vector of the helix, which we will take to point
in the $z$-direction, ${\bm q} = q\,{\hat{\bm z}}$, and $\lambda =
\Gamma_{\text{t}}\,m_0$ is the Stoner gap, with $m_0$ the magnetization
amplitude. $\tilde S_0$ in Eq.\ (\ref{eq:2.2a}) contains the action for
non-interacting band electrons plus, possibly, an interaction in the
spin-singlet channel. Finally, fluctuations of the helimagnetic order are taken
into account by generalizing ${\bm H}_0$ to a fluctuating classical field ${\bm
H}(x) = \Gamma_{\text{t}} {\bm M}(x) = {\bm H}_0 + \Gamma_{\text{t}} \delta{\bm
M}(x)$, where ${\bm M}(x)$ represents the spin density averaged over the
quantum mechanical degrees of freedom. $\chi_{\text{s}}^{ij}(x,y) =
\langle\delta M_i(x)\,\delta M_j(y)\rangle$ in Eq. (\ref{eq:2.1}) is the
magnetic susceptibility in the helimagnetic state, and the action
(\ref{eq:2.1}) has been obtained by integrating out the fluctuations
$\delta{\bm M}$.

The susceptibility $\chi_{\text{s}}$ was calculated before, see Sec. IV.E in I.
The part of $\chi_{\text{s}}$ that gives the dominant low-temperature
contributions to the various thermodynamic and transport quantities is the
helimagnon or Goldstone mode contribution. In I it was shown that the
helimagnon is a propagating mode with a qualitatively anisotropic dispersion
relation. For the geometry given above, the helimagnon is given in terms of
magnetization fluctuations that can be parameterized by (see (I.3.4a))
\bse
\label{eqs:2.3}
\bea
\delta M_x(x) &=& -m_0\, \phi(x)\,\sin qz
\label{eq:2.3a} \\
\delta M_y(x) &=& m_{0}\, \phi(x)\,\cos qz .
\label{eq:2.3b}
\eea
\ese
$\delta M_z = 0$ in an approximation that suffices to determine the leading
behavior of observables. In Eqs.\ (\ref{eqs:2.3}), $\phi$ is a phase variable.
In Fourier space, the phase-phase correlation function in the long-wavelength
and low-frequency limit is,
\bse
\label{eqs:2.4}
\be
\chi(k) \equiv \left\langle \phi(k)\,\phi (-k)\right\rangle = \frac{1}{2\NF}\,
\frac{q^2}{3\kF^2}\, \frac{1}{\omega_0^2({\bm k})-(i\Omega)^2}\ ,
\label{eq:2.4a}
\ee
with $\NF$ the electronic density of states per spin at the Fermi surface,
$\kF$ the Fermi wave number,\cite{F_footnote} $i\Omega \equiv i\Omega_n = i2\pi
Tn$ ($n=0,\pm 1,\pm2$, etc.) a bosonic Matsubara frequency, and $k = ({\bm
k},i\Omega)$. If we write ${\bm k} = ({\bm k}_{\perp},k_z)$, with ${\bm
k}_{\perp} = (k_x,k_y)$, the pole frequency is
\be
\omega_0({\bm k}) = \sqrt{c_z k_z^2 + c_{\perp} {\bm k}_{\perp}^4} .
\label{eq:2.4b}
\ee
Note the anisotropic nature of this dispersion relation, which implies that
$k_z$ scales as $k_{\perp}^2$, which in turn scales as the frequency or
temperature, $k_z \sim k_{\perp}^2 \sim T$.\cite{rotational_symmetry_footnote}
This feature will play a fundamental role in our explicit calculations in Sec.
\ref{sec:III} that relate 3D helimagnetic metals to 2D nonmagnetic metals, at
least in the ballistic limit. In a weak-coupling calculation the elastic
constants $c_z$ and $c_{\perp}$ are given by, see Eqs.\ (II.3.8),
\bea
c_z &=& \lambda^2 q^2/36 \kF^4
\nonumber\\
c_{\perp} &=& \lambda^2/96 \kF^4.
\label{eq:2.4c}
\eea
\ese

This specifies the action given in Eq.\ (\ref{eq:2.1}). In Fourier space, and
neglecting any spin-singlet interaction, it can be written,
\bse
\label{eq:2.5}
\begin{widetext}
\bea
S_{\text{eff}}[\bar\psi,\psi] &=& S_{0}[\bar\psi,\psi] +
S_{\text{int}}[\bar\psi,\psi],
\label{eq:2.5a} \\
S_{0}[\bar\psi,\psi] &=& \sum_{p} (i\omega - \xi_{\bm p}) \sum_{\sigma}
\bar\psi_{\sigma}(p)\,\psi_{\sigma}(p) + \lambda\, \sum_p
\left[\bar\psi_{\uparrow}(p)\,\psi_{\downarrow}(p+q) +
\bar\psi_{\downarrow}(p)\,\psi_{\uparrow }(p-q)\right],
\label{eq:2.5b} \\
S_{\text{int}}[\bar\psi,\psi] &=&-\frac{\lambda^2}{2}\frac{T}{V}\sum_{k}
\chi(k)\, \left[\delta n_{\uparrow\downarrow}(k-q) - \delta
n_{\downarrow\uparrow}(k+q)\right]\, \left[\delta n_{\uparrow\downarrow}(-k-q)
- \delta n_{\downarrow\uparrow}(-k+q)\right],
\label{eq:2.5c}
\eea
\end{widetext}
where $V$ is the system volume, and $i\omega \equiv i\omega_n = i 2\pi T
(n+1/2)$ ($n=0,\pm 1, \pm 2, \ldots)$ is a fermionic Matsubara frequency,
\be
n_{\sigma_1\sigma_2}(k) = \sum_p \bar\psi_{\sigma_1}(p)\,\psi_{\sigma_2}(p-k),
\label{eq:2.5d}
\ee
and
\be
\delta n_{\sigma_1\sigma_2}(k) = n_{\sigma_1\sigma_2}(k) - \langle
n_{\sigma_1\sigma_2}(k) \rangle.
\label{eq:2.5e}
\ee
\ese
Here $p = ({\bm p},i\omega)$, and $q$ denotes the four-vector $({\bm q},0)$.
Elsewhere in this paper we use the notation $q = \vert{\bm q}\vert$, which
should not lead to any confusion. In Eq.\ (\ref{eq:2.5b}), $\xi_{\bm p} =
\epsilon_{\bm p} - \epsilon_{\text{F}}$, with $\epsilon_{\text{F}}$ the Fermi
energy, and $\epsilon_{\bm p}$ the single-particle energy-momentum relation.
The latter we will specify in Eq.\ (\ref{eq:2.16}) below.

In the above effective action, $S_0$ represents noninteracting electrons on the
background of helimagnetic order that has been taken into account in a
mean-field or Stoner approximation. Fluctuations of the helimagnetic order lead
to an effective interaction between the electrons via an exchange of
helimagnetic fluctuations or helimagnons. This is reflected by the term
$S_{\text{int}}$, and the effective potential is proportional to the
susceptibility $\chi$.

\subsection{Canonical transformation to quasiparticle variables}
\label{subsec:II.B}

The action $S_0$ in Eq.\ (\ref{eq:2.5b}) above is not diagonal in either the
spin index or the wave number. A cursory inspection shows that by a suitable
combination of the fermion fields it is possible to diagonalize $S_0$ in spin
space. It is much less obvious that it is possible to find a transformation
that simultaneously diagonalizes $S_0$ in wave number space. In what follows we
construct such a transformation, i.e. we map the electronic helimagnon problem
onto an equivalent problem in which space is homogeneous.

Let us tentatively define a canonical transformation of the electronic
Grassmann fields $\bar\psi$ and $\psi$ to new quasiparticle fields
$\bar\varphi$ and $\varphi$, which also are Grassmann-valued, by
\bse
\label{eqs:2.6}
\bea
\bar\psi_{\uparrow}(p) &=& \bar\varphi_{\uparrow}(p) + \alpha^*_p\,
\bar\varphi_{\downarrow}(p)
\label{eq:2.6a} \\
\bar\psi_{\downarrow}(p) &=& \bar\varphi_{\downarrow}(p-q) + \beta^*_p\,
\bar\varphi_{\uparrow}(p-q)
\label{eq:2.6b} \\
\psi_{\uparrow}(p) &=& \varphi_{\uparrow}(p) + \alpha_p\,
\varphi_{\downarrow}(p)
\label{eq:2.6c} \\
\psi_{\downarrow}(p) &=& \varphi_{\downarrow}(p-q) + \beta_p\,
\varphi_{\uparrow}(p-q).
\label{eq:2.6d}
\eea
\ese
The coefficients $\alpha$ and $\beta$ are determined by inserting the Eqs.\
(\ref{eqs:2.6}) into Eq.\ (\ref{eq:2.5b}) and requiring this noninteracting
part of that action to be diagonal in the spin labels. This requirement can be
fulfilled by choosing them to be real and frequency independent, and given by
\bea
\alpha_p &=& \alpha_p^* = -\beta_{p+q} \equiv \alpha_{\bm p}
\nonumber\\
&=& \frac{1}{2\lambda}\left[\xi_{{\bm p}+{\bm q}} - \xi_{\bm p} +
\sqrt{(\xi_{{\bm p}+{\bm q}} - \xi_{\bm p})^2 + 4\lambda^2}\right].
\nonumber\\
\label{eq:2.7}
\eea
The noninteracting part of the action in terms of these new Grassmann fields is
readily seen to be diagonal in both spin space and wave number space.

To fully take into account the effect of the change of variables from the
fields $\bar\psi(p)$ and $\psi(p)$ to the fields $\bar\varphi(p)$ and
$\varphi(p)$ we also need to consider the functional integration that obtains
the partition function $Z$ from the action via
\bse
\label{eqs:2.8}
\be
Z = \int D[\bar\psi,\psi]\, e^{S_{\text{eff}}[\bar\psi,\psi]}.
\label{eq:2.8a}
\ee
The transformation of variables changes the integration measure as follows:
\bea
D[\bar\psi,\psi] &\equiv& \prod_{p,\sigma}\,d\bar\psi_{\sigma}(p)\,
d\psi_{\sigma}(p)
\nonumber\\
&=& \prod_{p,\sigma}\, J(p)\, d\bar\varphi_{\sigma}(p)\, d\varphi_{\sigma}(p),
\label{eq:2.8b}
\eea
with a Jabobian
\be
J(p) = (1 + \alpha_{\bm p}^2)^2.
\label{eq:2.8c}
\ee
\ese
We can thus normalize the transformation by defining final quasiparticle
variables $\bar\eta$ and $\eta$ by
\bse
\label{eqs:2.9}
\bea
\bar\psi_{\uparrow}(p) &=& \left[\bar\eta_{\uparrow}(p) + \alpha_{\bm p}\,
\bar\eta_{\downarrow}(p)\right]/\sqrt{1+\alpha_{\bm p}^2}\ ,
\label{eq:2.9a}\\
\bar\psi_{\downarrow}(p) &=& \left[\bar\eta_{\downarrow}(p-q) - \alpha_{{\bm
p}-{\bm q}}\,
\bar\eta_{\uparrow}(p-q)\right]/\sqrt{1+\alpha_{{\bm p}-{\bm q}}^2}\ , \nonumber\\
\label{eq:2.9b}\\
\psi_{\uparrow}(p) &=& \left[\eta_{\uparrow}(p) + \alpha_{\bm p}\,
\eta_{\downarrow}(p)\right]/\sqrt{1+\alpha_{\bm p}^2}\ ,
\label{eq:2.9c}\\
\psi_{\downarrow}(p) &=& \left[\eta_{\downarrow}(p-q) - \alpha_{{\bm p}-{\bm
q}}\,
\eta_{\uparrow}(p-q)\right]/\sqrt{1+\alpha_{{\bm p}-{\bm q}}^2}\ . \nonumber\\
\label{eq:2.9d}
\eea
\ese
In terms of these new Grassmann fields the Jacobian is unity, and the
noninteracting part of the action reads
\bse
\label{eqs:2.10}
\be
S_0[\bar\eta,\eta] = \sum_{p,\sigma}[i\omega - \omega_{\sigma}({\bm p})]\,
\bar\eta_{\sigma}(p)\eta_{\sigma}(p).
\label{eq:2.10a}
\ee
Here $\sigma = (\uparrow,\downarrow) \equiv (1,2)$, and
\be
\omega_{1,2}(\bm p) = \frac{1}{2}\left[\xi_{{\bm p}+{\bm q}} + \xi_{\bm p} \pm
\sqrt{(\xi_{{\bm p}+{\bm q}} - \xi_{\bm p})^2 + 4\lambda^2}\right].
\label{eq:2.10b}
\ee
The noninteracting quasiparticle Green function thus is
\be
G_{0,\sigma}(p) = \frac{1}{i\omega - \omega_{\sigma}({\bm p})}\ .
\label{eq:2.10c}
\ee
\ese
Physically, the Eqs.\ (\ref{eqs:2.10}) represent soft fermionic excitations
about the two Fermi surfaces that result from the helimagnetism splitting the
original band. The resonance frequencies $\omega_{1,2}$ are the same as those
obtained in II, see Eq.\ (II.3.19). We stress again that this Gaussian action
is diagonal in wave number space, so the quasiparticle system is homogeneous.

The interacting part of the action consist of two pieces. One contains terms
that couple the two Fermi surfaces. Because there is an energy gap, namely, the
Stoner gap $\lambda$, between these surfaces, these terms always lead to
exponentially small contributions to the electronic properties at low
temperatures, and will be neglected here. The second piece is, in terms of the
quasiparticle fields,
\bse
\label{eqs:2.11}
\be
S_{\text{int}}[\bar\eta,\eta] = -\frac{\lambda^2 q^2}{8\me^2} \frac{T}{V}
\sum_k \chi(k)\ \delta\rho(k)\, \delta\rho(-k).
\label{eq:2.11a}
\ee
Here we have defined
\be
\rho(k) = \sum_p \gamma({\bm k},{\bm p})\,\sum_{\sigma} \bar\eta_{\sigma}(p)\,
\eta_{\sigma}(p-k),
\label{eq:2.11b}
\ee
with
\be
\gamma({\bm k},{\bm p}) = \frac{2\me}{q}\,\frac{\alpha_{\bm p} - \alpha_{{\bm
p}-{\bm k}}}{\sqrt{1+\alpha_{\bm p}^2}\,\sqrt{1+\alpha_{{\bm p}-{\bm k}}^2}}\ ,
\label{eq:2.11c}
\ee
where $\me$ is the electron effective mass, and
\be
\delta\rho_{\sigma}(k) = \rho_{\sigma}(k) - \langle\rho_{\sigma}(k)\rangle.
\label{eq:2.11d}
\ee
\ese
An important feature of this result is the vertex function $\gamma({\bm k},{\bm
p})$, which is proportional to ${\bm k}$ for ${\bm k} \to 0$. The physical
significance is that $\phi$ is a phase, and hence only the gradient of $\phi$
is physically meaningful. Therefore, the $\phi$-susceptibility $\chi$ must
occur with a gradient squared in Eq.\ (\ref{eq:2.11a}). In the formalism of II
this feature became apparent only after complicated cancellations; in the
current formalism it is automatically built in. Also note the wave number
structure of the fermion fields in Eq.\ (\ref{eq:2.11b}), it the same as in a
homogeneous problem.

\subsection{Nonmagnetic disorder}
\label{subsec:II.C}

In the presence of nonmagnetic disorder there is an additional term in the
action. In terms of the original Grassmann variables, it reads
\be
S_{\text{dis}}[\bar\psi,\psi] = \int dx\, u({\bm x})\,\sum_{\sigma}
\bar\psi_{\sigma }(x)\psi_{\sigma}(x).
\label{eq:2.12}
\ee
Here $u({\bm x})$ is a random potential that we assume to be governed by a
Gaussian distribution with a variance given by
\be
\{u({\bm x})\,u({\bm y})\}_{\text{dis}} = \frac{1}{2\pi\NF\tau}\ \delta({\bm
x}-{\bm y}).
\label{eq:2.13}
\ee
Here $\{\ldots\}_{\text{dis}}$ denotes an average with respect to the Gaussian
probability distribution function, and $\tau$ is the (bare) elastic mean-free
time. Inserting the Eqs.\ (\ref{eqs:2.9}) into Eq. (\ref{eq:2.12}) yields
$S_{\text{dis}}[\bar\eta,\eta]$. Ignoring terms that couple the two Fermi
surfaces (which lead to exponentially small effects at low temperatures) yields
\bea
S_{\text{dis}}[\bar\eta,\eta] &=& \sum_{{\bm k},{\bm p}}\sum_{i\omega}
\sum_{\sigma}\frac{1+\alpha_{\bm k} \alpha_{\bm p}}{\sqrt{(1+\alpha_{\bm
k}^2)(1+\alpha_{\bm p}^2)}}\, u({\bm
k}-{\bm p})\, \nonumber\\
&&\hskip 30pt \times \bar\eta_{\sigma}({\bm k},i\omega)\, \eta_{\sigma}({\bm
p},i\omega).
\label{eq:2.14}
\eea

\subsection{Explicit quasiparticle action}

So far we have been very general in our discussion. In order to perform
explicit calculations, we need to specify certain aspects of our model. First
of all, we make the following simplification. In most of our calculations below
we will work in the limit where $\lambda \gg \vF q =
2\epsilon_{\text{F}}\,q/\kF$ with $\vF$ the Fermi velocity; i.e., the Stoner
splitting of the Fermi surfaces is large compared to the Fermi energy times the
ratio of the pitch wave number to the Fermi momentum. Since the dominant
contributions to the observables will come from wave vectors on the Fermi
surface, this implies that we can replace the transformation coefficients
$\alpha_{\bm p}$, Eq.\ (\ref{eq:2.7}), by unity in Eq.\ (\ref{eq:2.14}), and in
the denominator of Eq.\ (\ref{eq:2.11c}). In particular, this means that the
disorder potential in Eq.\ (\ref{eq:2.14}) couples to the quasiparticle
density:
\be
S_{\text{dis}}[\bar\eta,\eta] = \sum_{{\bm k},{\bm p}} u({\bm k}-{\bm p})
\sum_{i\omega} \sum_{\sigma} \bar\eta_{\sigma}({\bm k},i\omega)\,
\eta_{\sigma}({\bm p},i\omega).
\label{eq:2.15}
\ee

Second, we must specify the electronic energy-momentum relation $\epsilon_{\bm
p}\,$. For reasons already discussed in II, many of the electronic effects in
metallic helimagnets are stronger when the underlying lattice and the resulting
anisotropic energy-momentum relation is taken into account, as opposed to
working within a nearly-free electron model. We will assume a cubic lattice, as
appropriate for MnSi, so any terms consistent with cubic symmetry are allowed.
To quartic order in ${\bm p}$ the most general $\epsilon_{\bm p}$ consistent
with a cubic symmetry can be written
\be
\epsilon_{\bm p} = \frac{{\bm p}^2}{2\me} + \frac{\nu}{2\me\kF^2}(p_x^2 p_y^2 +
p_y^2 p_z^2 + p_z^2 p_x^2),
\label{eq:2.16}
\ee
with $\nu $ a dimensionless measure of deviations from a nearly-free electron
model. Generically one expects $\nu = O(1)$.

\begin{figure}[t,b,h]
\vskip -0mm
\includegraphics[width=8.0cm]{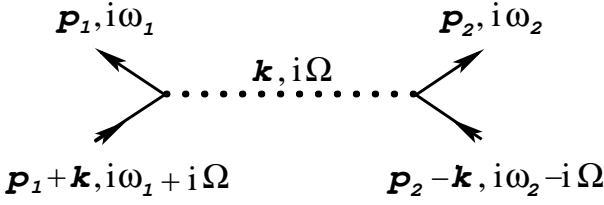}
\caption{The effective quasiparticle interaction due to helimagnons. Note that
the vertices depend on the quasiparticle momenta in addition to the helimagnon
momentum.}
\label{fig:1}
\end{figure}

With this model, and assuming $\lambda \gg q\vF$, which is typically satisfied,
given the weakness of the spin-orbit interaction, we obtain for the interaction
part of the action from Eqs.\ (\ref{eqs:2.11})
\begin{widetext}
\bea
S_{\text{int}} &=& \frac{-T}{V} \sum_{k,p_1,p_2} V(k;{\bm p_1},{\bm p_2})\,
\sum_{\sigma_1}\,\bigl[ \bar\eta_{\sigma_1}(p_1+k)\, \eta_{\sigma_1}(p_1) -
\left\langle \bar\eta_{\sigma_1}(p_1+k)\, \eta_{\sigma_1}(p_1)\right\rangle
\bigr]\,
\nonumber\\
&&\hskip 120pt \times\sum_{\sigma_2}\bigl[ \bar\eta_{\sigma_2}(p_2-k)
\eta_{\sigma_2}(p_2) - \left\langle \bar\eta_{\sigma_2}(p_2-k)
\eta_{\sigma_2}(p_2)\right\rangle\bigr],
\label{eq:2.17}
\eea
\end{widetext}
where the effective potential is
\bse
\label{eqs:2.18}
\be
V(k;{\bm p}_1,{\bm p}_2) = V_0\, \chi(k)\, \gamma({\bm k},{\bm p}_1)\, \gamma
(-{\bm k},{\bm p}_2).
\label{eq:2.18a}
\ee
Here,
\begin{equation}
V_0 = \lambda^2 q^2/8\me^2,
\label{eq:2.18b}
\end{equation}
and,
\be
\gamma ({\bm k},{\bm p}) = \frac{1}{2\lambda} \left[k_z + \frac{\nu}{\kF^2}\,
\bigl(k_z {\bm p}_{\perp}^2 + 2({\bm k}_{\perp}\cdot{\bm p}_{\perp}) p_z
\bigr)\right] + O(k^2).
\label{eq:2.18c}
\ee
\ese
The effective interaction is depicted graphically in Fig.\ \ref{fig:1}.

Examining the Eqs.\ (\ref{eq:2.17}) and (\ref{eqs:2.18}) we see three important
features. First, the effective potential is indeed proportional to ${\bm
k}^2\chi(k)$. As was mentioned after Eq.\ (\ref{eq:2.11d}), this is required
for a phase fluctuation effect. Second, the presence of the lattice, as
reflected by the term proportional to $\nu$ in Eq.\ (\ref{eq:2.18c}), allows
for a term proportional to $k_{\perp}^2 \chi(k)$ in the potential, which by
power counting is large compared to $k_z^2 \chi$, for reasons pointed out in
the context of Eq.\ (\ref{eq:2.4b}). It is this part of the potential that
results in the leading, and most interesting, low-temperature effects that will
be discussed in the next section of this paper, and in paper IV. Also as a
result of this feature, the dominant interaction between the quasiparticles is
not a density interaction, but rather an interaction between stress
fluctuations, due to the bilinear dependence on ${\bm p}$ of the dominant term
in $\gamma({\bm k},{\bm p})$. Third, the effective interaction is long-ranged,
due to the singular nature of the susceptibility $\chi(k)$ at long wave lengths
and low frequencies, see Eqs.\ (\ref{eqs:2.4}). This is a consequence of the
soft mode, the helimagnon, that mediates the interaction.

\begin{figure*}[t]
\vskip -0mm
\includegraphics[width=16.0cm]{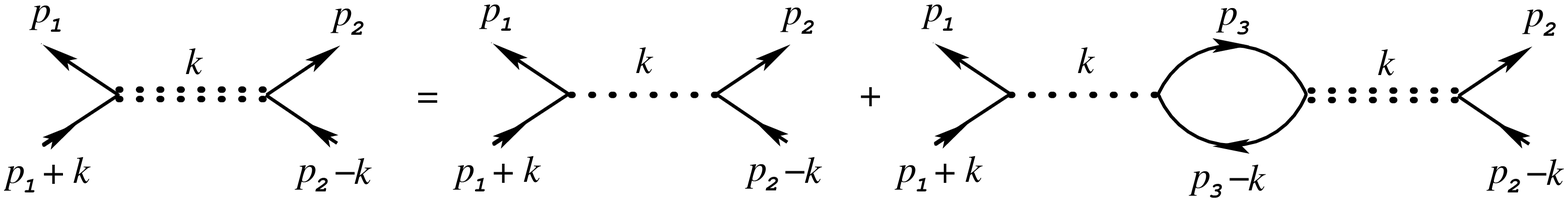}
\caption{Screening of the effective quasiparticle interaction.}
\label{fig:2}
\end{figure*}

In summary, we now have the following quasiparticle action:
\bse
\label{eqs:2.19}
\be
S_{\text{QP}}[\bar\eta,\eta] = S_0[\bar\eta,\eta] +
S_{\text{int}}[\bar\eta,\eta] + S_{\text{dis}}[\bar\eta,\eta],
\label{eq:2.19a}
\ee
with $S_0$ from Eqs.\ (\ref{eqs:2.10}), $S_{\text{int}}$ from Eqs.\
(\ref{eq:2.17}, \ref{eqs:2.18}), and $S_{\text{dis}}$ given by Eq.\
(\ref{eq:2.15}). The partition function is given by
\be
Z = \int D[\bar\eta,\eta]\, e^{S_{\text{QP}}[\bar\eta,\eta]},
\label{eq:2.19b}
\ee
with a canonical measure
\be
D[\bar\eta,\eta] = \prod_{p,\sigma}\, d\bar\eta_{\sigma}(p)\,
d\eta_{\sigma}(p).
\label{eq:2.19c}
\ee
\ese

\subsection{Screening of the quasiparticle interaction}
\label{subsec:II.E}

The quasiparticle interaction potential shown in Eqs.\ (\ref{eqs:2.18}) and
Fig.\ \ref{fig:1} must be screened, and an important question is whether this
will change its long-ranged nature. In the usual ladder or random-phase
approximation the screened potential $V_{\text{sc}}$ is determined by an
integral equation that is shown graphically in Fig.\ \ref{fig:2}, and
analytically given by
\bea
V_{\text{sc}}(k;{\bm p}_1,{\bm p}_2) &=& V(k;{\bm p}_1,{\bm p}_2) + \frac{T}{V}
\sum_{p_3} V(k;{\bm p}_1,{\bm p}_3) \nonumber\\
&&\hskip -40pt \times \sum_{\sigma} G_{0,\sigma}(p_3-k)\, G_{0,\sigma}(p_3)\,
V_{\text{sc}}(k;{\bm p}_3,{\bm p}_2). \nonumber\\
\label{eq:2.20}
\eea
It is convenient to define a screening factor $f_{\text{sc}}$ by writing
\be
V_{\text{sc}}(k;{\bm p}_1,{\bm p}_2) = V(k;{\bm p}_1,{\bm p}_2)\,
f_{\text{sc}}(k;{\bm p}_1,{\bm p}_2).
\label{eq:2.21}
\ee
Inserting Eq.\ (\ref{eq:2.21}) in Eq.\ (\ref{eq:2.20}) leads to an algebraic
equation for $f_{\text{sc}}$ with a solution
\bse
\label{eqs:2.22}
\be
f_{\text{sc}}(k;{\bm p}_1,{\bm p}_2) = \frac{1}{1 - V_0 \frac{1}{V} \sum_{\bm
p} \gamma({\bm k},{\bm p})\, \gamma(-{\bm k},{\bm p})\, \chi_{\text{L}}({\bm
p},i\Omega)},
\label{eq:2.22a}
\ee
where
\be
\chi_{\text{L}}({\bm p},i\Omega) = -T\sum_{i\omega} \sum_{\sigma}
G_{0,\sigma}({\bm p},i\omega)\,G_{0,\sigma}({\bm p},i\omega -i\Omega).
\label{eq:2.22b}
\ee
\ese
The most interesting effect of the screening is at $k\to 0$, and therefore we
need to consider only $\chi_{\text{L}}({\bm p},i\Omega = i0) \equiv
\chi_{\text{L}}({\bm p})$. This is essentially the Lindhard function, and we
use the approximation $(1/V)\sum_{\bm p} \vert{\bm p}\vert^n\,
\chi_{\text{L}}({\bm p}) \approx \kF^n\,\NF$. Neglecting prefactors of $O(1)$
this yields
\be
\frac{1}{V} \sum_{\bm p} \gamma({\bm k},{\bm p})\, \gamma(-{\bm k},{\bm p})\,
\chi_{\text{L}}({\bm p}) \approx \frac{\NF}{4\lambda^2}\,\left[(1+\nu)^2 k_z^2
+ \nu^2{\bm k}_{\perp}^2\right]. \nonumber
\ee
We finally obtain
\bse
\label{eqs:2.23}
\be
V_{\text{sc}}(k;{\bm p}_1,{\bm p}_2) = V_0\, \chi_{\text{sc}}(k)\, \gamma({\bm
k},{\bm p}_1)\, \gamma (-{\bm k},{\bm p}_2),
\label{eq:2.23a}
\ee
where
\be
\chi_{\text{sc}}(k) = \frac{1}{2\NF}\, \frac{q^2}{3\kF^2}\,
\frac{1}{\tilde\omega_0^2({\bm k})-(i\Omega)^2}\ .
\label{eq:2.23b}
\ee
Here
\be
\tilde\omega_0^2({\bm k}) = {\tilde c}_z\,k_z^2 + V_0\,\frac{\nu^2}{24}\,
\frac{q^2}{\kF^2\lambda^2}\,{\bm k}_{\perp}^2 + c_{\perp}\,{\bm k}_{\perp}^4,
\label{eq:2.23c}
\ee
with
\be
{\tilde c}_z = c_z\, \left[1 + \frac{q^2}{\kF^2}\,\left(
\frac{\epsilon_{\text{F}}}{\lambda}\right)^2 \right]\ .
\label{eq:2.23d}
\ee
\ese

We see that the screening has two effects on the frequency $\tilde\omega_0$
that enters the screened potential instead of the helimagnon frequency
$\omega_0$. First, it renormalizes the elastic constant $c_z$ by a term of
order $(q/\kF)^2\,(\epsilon_{\text{F}}/\lambda)^2$. This is a small effect as
long as $q\vF \ll \lambda$. Second, it leads to a term proportional to ${\bm
k}_{\perp}^2$ in $\tilde\omega_0^2$. Such a term also exists in the helimagnon
frequency proper, since the cubic lattice in conjunction with spin-orbit
effects breaks the rotational symmetry that is responsible for the absence of a
${\bm k}_{\perp}^2$ term in $\omega_0$, see Eq.\ (I.2.23) or (II.4.8), and it
is of order $b\, c_z q^2{\bm k}_{\perp}^2/\kF^2$, with $b=0(1)$. The complete
expression for $\tilde\omega_0^2$ is thus given by
\bse
\label{eqs:2.24}
\be
\tilde\omega_0^2({\bm k}) = {\tilde c}_z\,k_z^2 + {\tilde b}\,c_z
(q/\kF)^2\,{\bm k}_{\perp}^2 + c_{\perp}\,{\bm k}_{\perp}^4,
\label{eq:2.24a}
\ee
with
\be
{\tilde b} = b + (\epsilon_{\text{F}}/\lambda)^2.
\label{eq:2.24b}
\ee
\ese
As was shown in paper II, this puts a lower limit on the temperature range
where the isotropic helimagnon description is valid. In the absence of
screening, this lower limit is given by Eq.\ (II.4.9),
\bse
\label{eqs:2.25}
\be
T > T_{\text{so}} = b\lambda (q/\kF)^4.
\label{eq:2.25a}
\ee
This lower limit reflects the spin-orbit interaction effects that break the
rotational symmetry, and it is small of order $(q/\kF)^4$. Screening changes
this condition to
\be
T > {\tilde T}_{\text{so}} = {\tilde b}\lambda (q/\kF)^4,
\label{eq:2.25b}
\ee
\ese
which is still small provided $q\vF \ll \lambda$. We will therefore ignore the
screening in the remainder of this paper (as well as the spin-orbit term in
$\omega_0$), and return to a semi-quantitative discussion of its effects in
paper IV.

\section{Quasiparticle Properties}
\label{sec:III}

In this Section we use the effective quasiparticle action derived in Sec.
\ref{sec:II} to discuss the single-particle properties of an itinerant
helimagnet in the ordered phase. In Sec.\ \ref{subsec:III.A} we consider the
elastic scattering time in the helimagnetic state, in Sec.\ \ref{subsec:III.B}
we consider the effects of interactions on the single-particle relaxation rate
for both clean and disordered helimagnet, and in Sec.\ \ref{subsec:III.C} we
consider the effects of interactions on the single-particle density of states
for both clean and disordered helimagnets.

\subsection{Elastic relaxation time}
\label{subsec:III.A}

\begin{figure}[b]
\vskip -0mm
\includegraphics[width=4.5cm]{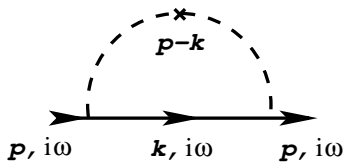}
\caption{Quasiparticle self energy due to quenched disorder. The directed solid
 line denotes the Green function, the dashed lines denote the disorder potential,
 and the cross denotes the disorder average.}
\label{fig:3}
\end{figure}

Helimagnetism modifies the elastic scattering rate, even in the absence of
interaction effects. To see this we calculate the quasiparticle self energy
from the action $S_0 + S_{\text{dis}}$ from Eqs.\ (\ref{eq:2.10a},
\ref{eq:2.14}). To first order in the disorder the relevant diagram is given in
Fig.\ \ref{fig:3}.
Analytically it is given by,
\be
\Sigma^{(3)}_{\sigma}({\bm p},i\omega) = \frac{-1}{8\pi\NF\tau}\, \frac{1}{V}
\sum_{\bm k}\, [1+\alpha_{\bm p}\,\alpha_{\bm k}]^2 G_{0,\sigma}({\bm
k},i\omega),
\label{eq:3.1}
\ee
with $G_{0}$ the noninteracting Green function from Eq.\ (\ref{eq:2.10c}). For
simplicity we put $\nu=0$ in Eq.\ (\ref{eq:2.16}), i.e., we consider nearly
free electrons. In the limit $q\vF \ll \lambda$ we obtain for the elastic
scattering rate, $1/\tau_{\text{el}} = -2\text{Im}\, \Sigma_{\sigma}({\bm
p},i0)$,
\bse
\label{eqs:3.2}
\be
\frac{1}{\tau_{\text{el}}} = \frac{1}{\tau}\,
\sqrt{1-\lambda/\epsilon_{\text{F}}}\ ,
\label{eq:3.2a}
\ee
In the opposite limit, $q\vF \gg \lambda$, we find
\be
\frac{1}{\tau_{\text{el}}} = \frac{1}{4\tau}\, \left[1- q/2\kF +
O((q/\kF)^2)\right]\ .
\label{eq:3.2b}
\ee
\ese

To first order in the disorder and to zeroth order in interactions, the
disorder averaged Green function is
\begin{equation}
G_{\sigma}(p) = \frac{1}{i\omega - \omega_{\sigma}({\bm p}) +
\frac{i}{2\tau_{\text{el}}}\, \sgn(\omega)}\ .
\label{eq:3.3}
\ee

\subsection{Interacting single-particle relaxation rate}
\label{subsec:III.B}

In this subsection we determine the single-particle relaxation rate due to
interactions, and its modification due to disorder in the ballistic limit.

\subsubsection{Clean helimagnets}
\label{subsubsec:III.B.1}

We first reproduce the results of paper II for the interaction-induced
single-particle relaxation rate. This serves as a check on our formalism, and
to demonstrate the technical ease of calculations within the quasiparticle
model compared to the formalism in papers I and II. To this end we calculate
the quasiparticle self energy for an action $S_0 + S_{\text{int}}$ from Eqs.\
(\ref{eq:2.10a}, \ref{eq:2.17}, \ref{eqs:2.18}). To first order in the
interaction there are two self-energy diagrams that are shown in Fig.\
\ref{fig:4}.
\begin{figure}[b]
\vskip -0mm
\includegraphics[width=8.0cm]{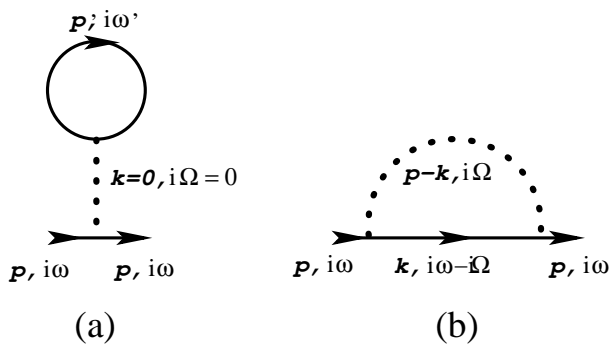}
\caption{Hartree (a), and Fock (b), contributions to the quasiparticle self
energy due to the effective interaction potential $V$ (dotted line).}
\label{fig:4}
\end{figure}
The direct or Hartree contribution, Fig.\ \ref{fig:4}(a), is purely real and
hence does not contribute to the scattering rate. The exchange or Fock
contribution, Fig.\ \ref{fig:4}(b), is given by
\be
\Sigma^{(4b)}_{\sigma}(p) = \frac{-T}{V} \sum_{k} V(k;{\bm p}-{\bm k},{\bm
p})\, G_{0,\sigma}(k-p).
\label{eq:3.4}
\ee

In order to compare with the results given in II, we consider the Fermi surface
given by $\omega_1({\bm p}) = 0$. The single-particle relaxation rate is given
by $1/\tau ({\bm k},\epsilon) = -2\text{Im}\Sigma_1({\bm k},\epsilon+i0)$. With
Eqs. (\ref{eq:2.10c}) and (\ref{eqs:2.18}) in Eq.\ (\ref{eq:3.4}), we find
\bea
\frac{1}{\tau({\bm k},\epsilon)} &=& 2 \int_{-\infty}^{\infty} du\
\left[n_{\text{B}}\left(\frac{u}{T}\right) + n_{\text{F}}\left(\frac{\epsilon
+u}{T}\right)\right]
\nonumber\\
&&\hskip -20pt\times V''({\bm p}-{\bm k};{\bm k},{\bm p};u)\, \delta(\epsilon +
u - \omega_1({\bm p})).
\label{eq:3.5}
\eea
Here $n_{\text{B}}(x) = 1/(e^x-1)$ and $n_{\text{F}}(x) = 1/(e^x+1)$ are the
Bose and Fermi distribution functions, respectively, and $V''({\bm k};{\bm
p}_1,{\bm p}_2;u) = \Im\,V(k=({\bm k},i\Omega\to u+i0);{\bm p}_1,{\bm p}_2)$ is
the spectrum of the potential. On the Fermi surface, $\epsilon =0$ and
$\omega_1({\bm k}) = 0$, we find for the relaxation rate $1/\tau({\bm k})
\equiv 1/\tau({\bm k},\epsilon=0)$,
\bse
\label{eqs:3.6}
\be
\frac{1}{\tau({\bm k})} = C_{\bm k}\, \frac{k_x^2 k_y^2 (k_x^2 -
k_y^2)^2}{(k_x^2 A_x^2 + k_y^2 A_y^2)^{3/2}}\,
\left(\frac{T}{\lambda}\right)^{3/2}.
\label{eq:3.6a}
\ee
The quantities $A_{x,y}$ and $C_{\bm k}$ are defined as
\be
A_{x,y} = 1 + \frac{\nu}{\kF^2}\,(k_{y,x}^2 + k_z^2),
\label{eq:3.6b}
\ee
and
\be
C_{\bm k} = \frac{B\nu^4}{8\lambda\kF^5}\, \frac{k_z^2}{\kF^2}\,
\frac{q^3\kF}{\me^2}\ ,
\label{eq:3.6c}
\ee
with
\be
B = \frac{48}{6^{1/4}} \int_0^{\infty} dx\,dz\, \frac{x^2}{\sqrt{z^2+x^4}}\,
\frac{1}{\sinh\sqrt{z^2+x^4}}\ .
\label{eq:3.6d}
\ee
\ese
They are identical with the objects defined in Eqs.\ (II.3.29), provided the
latter are evaluated to lowest order in $q/\kF$. The temperature dependence for
generic (i.e., $k_x \neq k_y$) directions in wave number space is thus
\be
\frac{1}{\tau({\bm k})} \propto \nu^4\lambda\, \left(\frac{q}{\kF}\right)^6\,
\left(\frac{\epsilon_{\text{F}}}{\lambda}\right)^2\,
\left(\frac{T}{T_q}\right)^{3/2},
\label{eq:3.7}
\ee
in agreement with Eq.\ (II.3.29d). $T_{q}$ is a temperature related to the
length scale where the helimagnon dispersion relation is valid, $\vert{\bm
k}\vert < q$. Explicitly, in a weakly coupling approximation, it is given by
\be
T_q = \lambda q^2/6\kF^2,
\label{eq:3.8}
\ee
see the definition after Eq.\ (II.3.9). $T_q$ also gives the energy or
frequency scale where the helimagnon crosses over to the usual ferromagnetic
magnon, see the discussion in Sec. IV.A of paper II.

The most interesting aspect of this result is that at low temperatures it is
stronger than the usual Fermi-liquid $T^2$ dependence, and nonanalytic in
$T^2$. Also note the strong angular dependence of the prefactor of the
$T^{3/2}$ in Eq.\ (\ref{eq:3.6a}). The experimental implications of this result
have been discussed in paper II.

\subsubsection{Disordered helimagnets in the ballistic limit}
\label{subsubsec:III.B.2}

We now consider effects to linear order in the quenched disorder. These can be
considered disorder corrections to the clean relaxation rate derived in the
previous subsection, or temperature corrections to the elastic relaxation rate.
The small parameter for the disorder expansion turns out to be
\be
\delta = 1/\sqrt{(\epsilon_{\text{F}}\tau)^2 T/\lambda} \ll 1.
\label{eq:3.8'}
\ee
That is, the results derived below are valid at weak disorder,
$\epsilon_{\text{F}}\tau \gg \sqrt{\lambda/T}$, or at intermediate temperature,
$T \gg \lambda/(\epsilon_{\text{F}}\tau)^2$. This can be seen from an
inspection of the relevant integrals in the disorder expansion, and will be
discussed in more detail in paper IV. For stronger disorder, or lower
temperature, the behavior of the quasiparticles is diffusive and will be
discussed elsewhere.\cite{us_tbp} The ballistic regime in a helimagnet is
different from that in a system of electrons interacting via a Coulomb
interaction, where the condition corresponding to Eq.\ (\ref{eq:3.8'}) reads
$T\tau \gg 1$.\cite{Zala_Narozhny_Aleiner_2001}

To first order in the disorder there are two types of diagrammatic
contributions to the single-particle relaxation rate: (A) diagrams that are
formally the same as those shown in Fig.\ \ref{fig:4}, except that the solid
lines represent the disorder-averaged Green function given by Eq.\
(\ref{eq:3.3}), and (B) diagrams that have one explicit impurity line. The
latter are shown in Fig.\ \ref{fig:5}. It is easy to show that the various
Hartree diagrams do not contribute. The class (A) Fock contribution to the self
energy is given by Eq.\ (\ref{eq:3.4}), with $G_{0,\sigma}$ replaced by
$G_{\sigma}$ from Eq.\ (\ref{eq:3.3}).
\begin{figure}[t]
\vskip -0mm
\includegraphics[width=8.0cm]{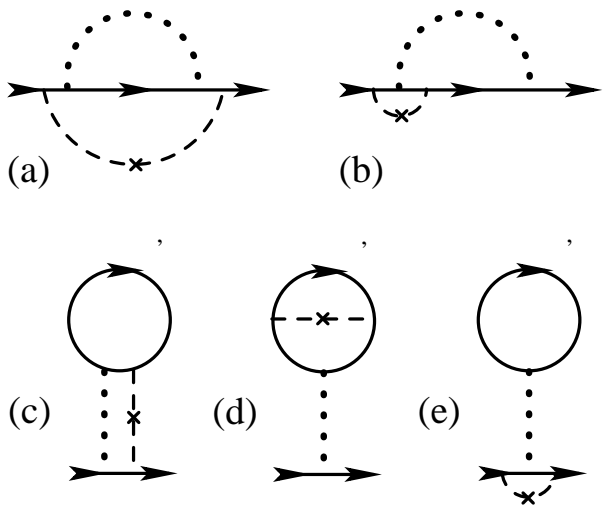}
\caption{Fock (a,b) and Hartree (c-e) ontributions to the self-energy in the
ballistic limit. See the text for additional information.}
\label{fig:5}
\end{figure}

Power counting shows that, (1) the leading contribution to the single-particle
relaxation rate in the ballistic limit is proportional to $T$, (2) the diagrams
of class (A) do not contribute to this leading term, and (3) of the diagrams of
class (B) only diagram (a) in Fig.\ \ref{fig:5} contributes. Analytically, the
contribution of this diagram to the self energy is
\bea
\Sigma^{(5a)}_{\sigma}({\bm p},i\omega) &\equiv&
\Sigma^{(5a)}_{\sigma}(i\omega) \nonumber\\
&&\hskip -40pt =\frac{-1}{2\pi\NF\tau} \frac{T}{V} \sum_{{\bm k},i\Omega}
\frac{1}{V}\sum_{{\bm p}'} V({\bm k},i\Omega;{\bm
p}'-{\bm k},{\bm p}')\, \nonumber\\
&&\hskip -20pt \times G_{\sigma}^2({\bm p}',i\omega)\, G_{\sigma}({\bm p}'-{\bm
k},i\omega - i\Omega).
\label{eq:3.9}
\eea
Notice that $\Sigma^{(5a)}$ does not depend on the wave vector. This leads to
the following leading disorder correction to the clean single-particle rate in
Eqs.\ (\ref{eqs:3.6}),
\bse
\label{eqs:3.10}
\bea
\delta(1/\tau({\bm p})) &\equiv& \delta(1/\tau) \nonumber\\
&&\hskip -40pt = \frac{V_0}{2\pi\NF\tau}\frac{1}{V}\sum_{\bm k}
\int_{-\infty}^{\infty} \frac{du}{\pi}\, n_{\text{F}}(u/T)\,\chi''({\bm k},u)\,
\nonumber\\
&&\hskip 50pt \times\text{Im}\,L^{++,-}({\bm k}).
\label{eq:3.10a}
\eea
Here $\chi''$ is the spectral function is the spectral function of the
susceptibility in Eq.\ (\ref{eq:2.4a}),
\bea
\chi''({\bm k},u) &=& \text{Im}\,\chi({\bm k},i\Omega \to u+i0)
\nonumber\\
&&\hskip -30pt =\frac{\pi}{12\NF}\, \frac{q^2}{\kF^2}\,\frac{1}{\omega_0({\bm
k})} \left[\delta(u - \omega_0({\bm k})) - \delta(u + \omega_0({\bm
k}))\right], \nonumber\\
\label{eq:3.10b}
\eea
and $L^{++,-}$ is an integral that will also appear in the calculation of the
conductivity in paper IV,
\bea
L^{++,-}({\bm k}) &=& \frac{1}{V}\sum_{\bm p} \gamma({\bm k},{\bm p})\,
\gamma({\bm k},{\bm p}-{\bm k})\,G_R^2({\bm p})\,G_A({\bm p}-{\bm k})
\nonumber\\
&=&i\nu^2\,\frac{2\pi}{3}\,\frac{\NF\me^2}{\lambda^2\kF^2} + O(1/\tau,{\bm
k}_{\perp}^2),
\label{eq:3.10c}
\eea
\ese
with $G_{R,A}({\bm p}) = G_1({\bm p},i\omega \to \pm i0)$ the retarded and
advanced Green functions.

Inserting Eqs.\ (\ref{eq:3.10b}, \ref{eq:3.10c}) into Eq.\ (\ref{eq:3.10a}) and
performing the integrals yields, for the leading temperature dependent
contribution to $\delta(1/\tau)$,
\be
\delta(1/\tau) = \frac{\nu^2\pi\ln 2}{12\sqrt{6}\,\tau}\,
\left(\frac{q}{\kF}\right)^5\, \frac{\epsilon_{\text{F}}}{\lambda}\,
\frac{T}{T_{q}}\ .
\label{eq:3.11}
\ee

Notice that $\delta(1/\tau)$ has none of the complicated angular dependence
seen in the clean relaxation rate, Eq.\ (\ref{eq:3.6a}). While quenched
disorder is expected to make the scattering process more isotropic in general,
it is quite remarkable that there is no angular dependence whatsoever in this
contribution to $\delta(1/\tau)$.

\subsection{The single-particle density of states}
\label{subsec:III.C}

The single-particle density of states, as a function of the temperature and the
energy distance $\epsilon$ from the Fermi surface, can be defined in terms of
the Green functions by\cite{Altshuler_Aronov_1984}
\be
N(\epsilon,T) = \frac{1}{\pi V}\sum_{\bm p}\sum_{\sigma}\text{Im}\,{\cal
G}_{\sigma}({\bm p},i\Omega\to \epsilon + i0).
\label{eq:3.12}
\end{equation}
Here ${\cal G}$ is the fully dressed Green function. The interaction correction
to $N$, to first order in the interaction, can be written
\be
\delta N(\epsilon) = \frac{-1}{\pi V}\sum_{\bm p}\sum_{\sigma}\text{Im}\,
\left[ G_{\sigma}^2({\bm p},i\Omega)\, \Sigma_{\sigma}({\bm
p},i\Omega)\right]_{i\Omega \to \epsilon + i0}\ ,
\label{eq:3.13}
\ee
with the dominant contribution to the self-energy $\Sigma$ given by Eq.\
(\ref{eq:3.9}). From the calculation in Sec.\ \ref{subsubsec:III.B.2} we know
that the leading contribution to $\Sigma$ is of order $T/\tau$, and the
integral over the Green functions is of $O(\tau\,T^0)$, so $\delta N$
potentially has a contribution of $O(\tau^0 T)$. However, an inspection of the
integrals shows that this term has a zero prefactor. Hence, to this order in
the interaction, there is no interesting contribution to the
temperature-dependent density of states.

\section{Discussion and Conclusion}
\label{sec:IV}

In summary, there have been two important results in this paper. First, we have
shown that there is a canonical transformation that diagonalizes the action for
helimagnets in the ordered state in spin space, and in the clean limit maps the
problem onto a homogenous fermion action. This transformation enormously
simplifies the calculations of electronic properties in an itinerant electron
system with long-ranged helimagnetic order. As was mentioned in the
Introduction, our model and conclusions are valid whether or not the
helimagnetism is due to the conduction electrons. We have also discussed the
effect of screening on the effective interaction that was first derived in
paper II. We have found that screening makes the interaction less long-ranged,
as is the case for a Coulomb potential. However, in contrast to the latter,
screening does not introduce a true mass in the effective electron-electron
interaction in a helimagnet. Rather, it removes the qualitative anisotropy
characteristic of the unscreened potential in a rotationally invariant model,
and introduces a term similar to one that is also generated by the spin-orbit
interaction in a lattice model.

We have used the transformed action to compute a number of the low-temperature
quasiparticle properties in a helimagnet. Some of the results derived here
reproduce previous results that were obtained with more cumbersome methods in
paper II. We then added quenched nonmagnetic disorder to the action, and
considered various single-particle observables in the ballistic limit. All of
these results are new. The second important result in this paper is our
calculation of the single-particle relaxation rate in systems with quenched
disorder in the ballistic limit, $\tau^2 T\epsilon_{\text{F}}^2/\lambda > 1$,
where we find a linear temperature dependence. This non-Fermi-liquid result is
to be contrasted with the previously derived $T^{3/2}$ leading term in clean
helimagnets, and the usual $T^2$ behavior in clean Fermi liquids.

In paper IV of this series we will treat the interesting problem of transport
in clean and weakly disordered electron systems with long-ranged helimagnetic
order. Specifically, we will use the canonical transformation introduced here
to compute the electrical conductivity. In the clean limit we will recover the
result derived previously in paper II, while in the ballistic regime we find a
leading temperature dependence proportional to $T$. This linear term is
directly related to the $T$-term found above for the single-particle relaxation
rate. For the case of the electrical conductivity, the $T$ term is much
stronger than either the Fermi liquid contribution ($T^2$) or the contribution
from the helimagnon scattering in the clean limit ($T^{5/2}$).

A detailed discussion of the experimental consequences of these results will be
given in paper IV. There we will also give a complete discussion of the
limitations of our results, and in particular of the various temperature scales
in the problem, including the one introduced by screening the effective
potential.

The linear temperature terms found here for the various relaxation times in
bulk helimagnets is closely related to the linear $T$ terms found in
two-dimensional nonmagnetic metals, also in a ballistic
limit.\cite{Zala_Narozhny_Aleiner_2001} The analogy between 3D helimagnets and
2D nonmagnetic materials is a consequence of the anisotropic dispersion
relation of the helical Goldstone mode or helimagnons. Technically, a typical
integral that appears in the bulk helimagnet case is of the form
\begin{eqnarray*}
\int dk_z \int d{\bm k}_{\perp}\,{\bm k}_{\perp}^2\ \delta(\Omega^2 - k_z^2 -
{\bm k}_{\perp}^4)\,f(k_z,{\bm k}_{\perp}) &\propto&
\nonumber\\
&& \hskip -150pt \int d{\bm k}_{\perp}\,{\bm
k}_{\perp}^2\,\frac{\Theta(\Omega^2-{\bm k}_{\perp}^4)}{\sqrt{\Omega^2 - {\bm
k}_{\perp}^4}}\,f(k_z=0,{\bm k}_{\perp}),
\end{eqnarray*}
and the dependence of $f$ on $k_z$ can be dropped since it does not contribute
to the leading temperature scaling. The prefactor of the ${\bm k}_{\perp}$
dependence of $f$ is of O(1) in a scaling sense. As a result, the 3D integral
over ${\bm k}$ behaves effectively like the integral in the 2D nonmagnetic
case. Physically the slow relaxation in the plane perpendicular to the pitch
vector makes the physics two-dimensional.

\acknowledgments

This research was supported by the National Science Foundation under Grant Nos.
DMR-05-30314 and DMR-05-29966.


\begin{thebibliography}{15}
\bibitem[{F_f()}]{F_footnote}
\bibinfo{note}{Due to the Stoner splitting, one strictly speaking has to
  distinguish between Fermi-surface properties, such as the Fermi wave number,
  the density of states at the Fermi surface, etc., in the two Stoner bands.
  For a weak helimagnet the differences between these quantities are small, and
  we will systematically neglect them. This amounts to working to lowest order
  in the small parameter $\lambda/\epsilon_{\text{F}}$.}
\bibitem[{rot()}]{rotational_symmetry_footnote}
\bibinfo{note}{Strictly speaking, this is true only in systems with rotational
  symmetry. In an actual helimagnet on a lattice, the spin-orbit interaction
  leads to a term proportional to $c_z\,(q/\kF)^2 {\bm k}_{\perp}^2$ in
  $\omega_0^2({\bm k})$. We will discuss this in the context of screening the
  phase-phase susceptibility in Sec.\ \ref{subsec:II.E} below.}
\end{thebibliography}

\end{document}